\documentstyle[12pt,epsf]{article}

\newcommand{\nc}{\newcommand}
\nc{\fdiag}{0}
\nc{\bg}{B. Grz${{{\rm a}_{}}_{}}_{\hskip -0.18cm\varsigma}$dkowski}
\nc{\lsp}{\;\;\;\;\;\;\;\;}
\nc{\beq}{\begin{equation}}   \nc{\eeq}{\end{equation}}
\nc{\bea}{\begin{eqnarray}}   \nc{\eea}{\end{eqnarray}}
\nc{\baa}{\begin{array}}      \nc{\eaa}{\end{array}}
\nc{\bit}{\begin{itemize}}    \nc{\eit}{\end{itemize}}
\nc{\ben}{\begin{enumerate}}  \nc{\een}{\end{enumerate}}
\nc{\bce}{\begin{center}}     \nc{\ece}{\end{center}}
\nc{\non}{\nonumber}
\nc{\lumun}{\;{\hbox {pb}^{-1}}{\hbox {yr}^{-1}}}
\nc{\hc}{\hbox {h.c.}}
\nc{\re}{\hbox {Re}}
\nc{\im}{\hbox {Im}}
\nc{\etal}{\hbox{et al.}}
\nc{\prdj}[3]{{\it Phys.\ Rev.}\ {{\bf D{#1}} (#2), #3}}
\nc{\prlj}[3]{{\it Phys.\ Rev.\ Lett.}\ {{\bf {#1}} (#2), #3}}
\nc{\plbj}[3]{{\it Phys.\ Lett.}\ {{\bf B{#1}} (#2), #3}}
\nc{\npbj}[3]{{\it Nucl.\ Phys.}\ {{\bf B{#1}} (#2), #3}}
\nc{\ptpj}[3]{{\it Prog.\ Theor.\ Phys.}\ {{\bf {#1}} (#2), #3}}
\nc{\zfpj}[3]{{\it Z.\ Phys.}\ {{\bf C{#1}} (#2), #3}}
\nc{\mplaj}[3]{{\it Mod.\ Phys.\ Lett.}\ {{\bf A{#1}} (#2), #3}}
\nc{\rmpj}[3]{{\it Rev.\ Mod.\ Phys.}\ {{\bf {#1}} (#2), #3}}
\nc{\ijmpaj}[3]{{\it Int.\ J.\ of\ Mod.\ Phys.}\ {{\bf A{#1}} (#2), #3}}
\nc{\ra} {\rightarrow}
\nc{\cw}{\cos\theta_W}        \nc{\sw}{\sin\theta_W}
\nc{\ttbar}{t\bar{t}}
\nc{\bbbar}{b\bar{b}}
\nc{\tanb} {\tan \beta}
\nc{\twbdec} {t\rightarrow W^+ b}
\nc{\tbwbdec} {\bar{t} \rightarrow W^- \bar{b}}
\nc{\hprod} {e^+e^- \ra Z^\ast \ra H Z}
\nc{\epem} {e^+e^-}
\nc{\wpwm} {W^+W^-}
\nc{\tbar} {\bar{t}}
\nc{\bbar} {\bar{b}}
\nc{\wpp} {W^+}
\nc{\mt}{m_t}
\nc{\mts}{m_t^2}
\nc{\mw} {m_W}
\nc{\mws} {m_W^2}
\nc{\mz} {m_Z}
\nc{\mzs} {m_Z^2}
\nc{\mh} {m_H}
\nc{\mhs} {m_H^2}
\nc{\ma} {m_A}
\nc{\mas} {m_A^2}
\nc{\hdec}{H \ra t\bar{t}}
\nc{\ttbardec}{\ttbar \ra W^+W^-\bbbar}
\nc{\po}{\Phi_1}
\nc{\pod}{\Phi_1^\dagger}
\nc{\pht}{\Phi_2}
\nc{\phtd}{\Phi_2^\dagger}
\nc{\phtt}{{\tilde{\Phi}}_2}
\nc{\popo}{\po^\dagger\po}
\nc{\phtpt}{\pht^\dagger\pht}
\nc{\popt}{\po^\dagger\pht}
\nc{\phtpo}{\pht^\dagger\po}
\nc{\sq}{\sqrt{2}}
\nc{\nsd} {N_{SD}}
\nc{\ntt} {N_{tt}}

\def\nmess{N_m}
\def\mbino{M_{\wtil B}}
\def\mplanck{M_{\rm Planck}}

\def\sq{\wt q}

\def\msq{m_{\sq}}

\def\slepl{\wt \ell_L}
\def\mslepl{m_{\slepl}}
\def\slepr{\wt \ell_R}
\def\mslepr{m_{\slepr}}

\def\susy{{\rm SUSY}}

\def\cnone{\wt\chi^0_1}

\def\slash#1{#1\hskip-8pt/\hskip5pt}

\def\etmiss{\slash E_T}
\def\etmin{\slash E_T^{\rm min}}

\def\sq{\wt q}

\def\msq{m_{\sq}}

\def\slepl{\wt \ell_L}
\def\mslepl{m_{\slepl}}
\def\slepr{\wt \ell_R}
\def\mslepr{m_{\slepr}}
\def\fbi{~{\rm fb}^{-1}}

\def\pbi{~{\rm pb}^{-1}}
\def\pb{~{\rm pb}}
\def\kev{~{\rm keV}}

\def\gev{~{\rm GeV}}
\def\tev{~{\rm TeV}}

\def\mt{m_t}

\def\mcnone{m_{\cnone}}

\def\wt{\widetilde}

\def\stauone{\wt \tau_1}

\def\wtil{\widetilde}
\def\what{\widehat}

\def\vev#1{\langle {#1}\rangle}

\def\lsim{\mathrel{\raise.3ex\hbox{$<$\kern-.75em\lower1ex\hbox{$\sim$}}}}
\def\gsim{\mathrel{\raise.3ex\hbox{$>$\kern-.75em\lower1ex\hbox{$\sim$}}}}

\def\pbi{~{\rm pb}^{-1}}
\def\fbi{~{\rm fb}^{-1}}

\def\pb{~{\rm pb}}

\def\gev{\,{\rm GeV}}
\def\tev{\,{\rm TeV}}

\def\wt{\widetilde}

\def\h{h}
\def\a{a}
\def\mh{m_{\h}}
\def\ma{m_{\a}}

\def\calo{{\cal O}}
\def\eg{{\it e.g.}}
\def\etal{{\it et al.}}
\def\epem{e^+e^-}

\def\lsim{\mathrel{\raise.3ex\hbox{$<$\kern-.75em\lower1ex\hbox{$\sim$}}}}
\def\gsim{\mathrel{\raise.3ex\hbox{$>$\kern-.75em\lower1ex\hbox{$\sim$}}}}
\def\@versim#1#2{\vcenter{\offinterlineskip
        \ialign{$\m@th#1\hfil##\hfil$\crcr#2\crcr\sim\crcr } }}

\def\slash#1{#1\hskip-8pt/\hskip4pt}

\def\etmiss{\slash E_T}

\def\ie{{\it i.e.}}

\def\gam{\gamma}

\def\nsd{N_{SD}}

\def\pbi{~{\rm pb}^{-1}}
\def\fbi{~{\rm fb}^{-1}}

\def\pb{~{\rm pb}}

\def\gev{\,{\rm GeV}}
\def\tev{\,{\rm TeV}}

\def\wt{\widetilde}

\def\sq{\wt q}

\def\msq{m_{\sq}}

\def\slepl{\wt \ell_L}
\def\mslepl{m_{\slepl}}
\def\slepr{\wt \ell_R}
\def\mslepr{m_{\slepr}}

\def\tanb{\tan\beta}

\def\mt{m_t}

\def\mz{m_Z}
\def\mw{m_W}

\def\cnone{\wt\chi^0_1}

\def\mcnone{m_{\cnone}}

\def\h{h}
\def\mh{m_{\h}}

\def\stauone{\wt \tau_1}

\begin{document}
%\begin{flushright}
%Version: \today\hfill
%\end{flushright}
%
\font\fortssbx=cmssbx10 scaled \magstep2
\medskip
$\vcenter{
\hbox{\fortssbx University of California - Davis}
}$
\hfill
$\vcenter{
\hbox{\bf UCD-97-15} 
%\hbox{\bf LBNL-40399}
\hbox{June, 1997}
}$
\vspace*{1cm}
\begin{center}
{\large{\bf Maximizing Hadron Collider Sensitivity to Gauge-Mediated
Supersymmetry Breaking Models}}\\
\rm
\vspace*{1cm}
{\bf  C.-H. Chen and J.F. Gunion}\\
\vspace*{1cm}
{\em Department of Physics, 
University of California, Davis, CA, USA }
%\vspace*{3cm}
\end{center}
\begin{abstract}
We consider typical hadron collider detector 
signals sensitive to delayed decays of the
lightest neutralino to photon plus goldstino and demonstrate
the potential for substantially 
increasing the portion of the general parameter space of a
gauge-mediated supersymmetry breaking model that can be probed at the Tevatron.
\end{abstract}
\vspace{.5in}

%\setcounter{page}{0}
%\thispagestyle{empty}
%\newpage

It has been conventional~\cite{ddrt,akkmm,dtw,bbct,bmpz}
to analyze the sensitivity of the CDF and D0
experiments to supersymmetric models with gauge-mediated 
supersymmetry breaking~\cite{basicgmsb} (GMSB models)
under the assumption that the lightest neutralino decays promptly
to a photon plus the goldstino.  In this case, the hadron collider
signal for supersymmetry
would be events containing two or more isolated photons, deriving
from decay of two or more neutralinos, plus missing energy
(and possibly other particles).
However, in the GMSB models
for which explicit constructions are available, the neutralino 
decay is most naturally characterized
by a substantial decay length, possibly of order tens to hundreds of meters.
Events in which one or more of the neutralinos travels
part way through the detector
and then decays can be a substantial fraction of the total,
as can events in which all the neutralinos exit the detector
before decaying (the latter leading to the jets 
plus missing energy and other conventional signals for supersymmetry).
In this Letter, we assess the increase in parameter space
coverage at the Tevatron that results by
combining the two-photon plus missing energy signal and the standard
jets plus missing energy signal with a typical signature for a delayed
neutralino decay within the detector. We find that the range
of parameter space that can be probed is substantially increased
for luminosities expected in Run II and at TeV33.

In the simpler GMSB models~\cite{basicgmsb}
the phenomenology is primarily determined by just two mass scales:
the scale $\sqrt{ \vev{F}}$
at which \susy-breaking occurs and a scale $\Lambda\equiv \vev{F_S}/\vev{S}$
arising in the so-called messenger sector that communicates \susy-breaking to
the usual supersymmetric (sparticle) 
partners of the Standard Model (SM) particles.
Here, $\vev{F}$ is the magnitude of the expectation value of a typical $F$-term
component in the sector of the theory responsible for supersymmetry breaking;
$F_S$ and $S$ are the scalar and $F$-term components of the singlet 
superfield, $\what S$, member of the messenger sector.
(In what follows, we will use $F$ and $F_S$ to denote $\vev{F}$
and $\vev{F_S}$.)
A third mass scale is $M$, the mass-scale of the fermion members of the
messenger sector. The scale $\sqrt F$ determines
the mass and couplings of the goldstone fermion, or goldstino ($G$),
resulting from spontaneous \susy-breaking~\cite{fayet,others}.
The scales $\Lambda$ and $M$ determine
the soft-supersymmetry-breaking masses
of the sfermions and gauginos.

For the SU(3), SU(2) and U(1) gauginos,
\begin{equation}
M_i(M)=k_i\nmess\, g\left({\Lambda\over M}\right) 
{\alpha_i(M)\over 4\pi}\Lambda\,,
\label{inomasses}
\end{equation}
where $k_3=k_2=1$, $k_1=5/3$, respectively.
For the squarks and sleptons 
\begin{equation}
{
%\textstyle
m_i^2(M)=2\Lambda^2\nmess f\left({\Lambda\over M}\right)
\left[c_3\left({\alpha_3(M)\over
4\pi}\right)^2+c_2\left({\alpha_2(M)\over 4\pi}\right)^2+{5\over 3}\left({Y\over
2}\right)^2\left({\alpha_1(M)\over 4\pi}\right)^2\right]\,,}
\label{scalarmasses}
\end{equation}
with $c_3=4/3$ (color triplets), $c_2=3/4$ (weak doublets), and $Y/2=Q-T_3$. 
The integer $\nmess$ (which must be $\leq 4$ to avoid Landau poles)
depends upon the (model-dependent) matter 
content of the messenger sector. To avoid negative mass-squared
for bosonic members of the messenger sector 
$M/\Lambda> 1$ is required; $M/\Lambda\geq 1.1$ is
preferred to avoid fine-tuning, for which $f(\Lambda/M)\simeq 1$ and
$1\leq g(\Lambda/M)\leq 1.23$. For $M/\Lambda\geq 2$, $1\leq g(\Lambda/M)\leq
1.045$. In this Letter, we will consider masses obtained with $g=f=1$, for
which Eqs.~(\ref{inomasses}) and (\ref{scalarmasses}) imply
$
\msq:\mslepl:\mslepr:M_1=11.6:2.5:1.1:\sqrt{\nmess}\,.
$

These results at scale $M$ must be evolved down to the scale of the
actual sparticle masses, denoted $Q$.
The resulting gaugino masses are given
by replacing $\alpha_i(M)$ in Eq.~(\ref{inomasses}) by $\alpha_i(Q)$,
a useful reference result being
$
\Lambda \sim({80\tev/ \nmess})
\left({M_1/ 100\gev}\right)\,.
$
Evolution of the sfermion masses is detailed in Ref.~\cite{martin}.
Most important is the ratio
$$\mslepr/M_1=\sqrt{6/5}\sqrt{r_1/\nmess-(5/33)(1-r_1)}\,,$$
where $r_1=(\alpha_1(M)/\alpha_1(Q))^2$. For the very broad
range of $Q\geq\mz$
and $M\leq 3\times 10^6\tev$, $1<r_1\lsim 1.5$, in which case
the lightest of the sparticle partners of the SM
particles is the $\wtil B$ for $\nmess=1$ or the $\slepr$ (more precisely,
the $\stauone$) for $\nmess\geq2$. Our numerical results are obtained
by evolving from $M=1.1\Lambda$. Aside from $\Lambda$, the only other
free parameters of the model are then 
$\tanb$ and sign($\mu$) (which, combined with
the known value of $\mz$, fix the low-energy $\mu$ and $B$ parameters
describing Higgs superfield and scalar field mixing). For this
study, we have taken $\tanb=2$ and sign($\mu)=-1$.

For the goldstino, one finds
$
m_G=F/ (\sqrt 3\mplanck)\sim 2.5
\left({\sqrt{F}/ 100\tev}\right)^2~{\rm eV}\,.
$
%In the usual minimal supergravity model, with supersymmetry
%breaking communicated via interactions at $\mplanck$, one has
%$\sqrt F\sim \sqrt{\mw\mplanck}\sim \calo\left(10^8 \tev\right)$, in which case 
%$m_G\sim \calo\left(1\tev\right)$; since the $G$
%is also very weakly interacting,
%it is then irrelevant to low-energy physics.  
In GMSB models, the
small $\sqrt F\sim \calo(100-1000\tev)$ values envisioned imply
that $G$ will be the lightest supersymmetric particle.
In this case, most early-universe scenarios lead to the requirement
$m_G\lsim 1\kev$ (equivalent to $\sqrt F\lsim 2000\tev$)
to avoid overclosing the universe
\cite{coslimit,cosmur}. The coupling of a sparticle to its SM partner plus
the $G$ is inversely proportional to $F$ and is
very weak for all $F$ values of interest.
As a result, all the sparticles other than the next-to-lightest super
particle (NLSP), \ie\ the $\cnone$ or $\slepr$, undergo
chain decay down to the NLSP. The NLSP finally decays to the $G$:
\eg\ $\wtil B\to \gam G$ (and $ZG$ if $\mbino>\mz$) 
or $\slepr\to \ell G$. The $c\tau$ for
NLSP decay depends on $\sqrt F$; \eg\ for $\nmess=1$
\begin{equation}
(c\tau)_{\cnone=\wtil B\to \gam G}\sim 130
\left({100\gev\over \mbino}\right)^5
\left({\sqrt F\over 100\tev}\right)^4\mu {\rm m}
\label{ctauform}
\end{equation}
If $\sqrt F\sim 2000\tev$ (the upper limit from cosmology),
then $c\tau\sim 21$m for $\mbino=100\gev$; 
$\sqrt F\sim 100\tev$ implies a short but vertexable decay length. 
Thus, the signatures for GMSB models are crucially dependent
on $\sqrt F$.

In the GMSB models~\cite{basicgmsb}, the \susy-breaking sector lies at
a higher mass scale than the messenger sector and
\susy-breaking is then communicated by a new strong interaction 
{\it at two-loops} to the messenger sector.
The two-loop communication
between the two sectors implies that $F/F_S\sim (16\pi^2/\alpha_m^2)
\geq 2.5\times 10^4$ for $g_m\leq 1$,
where $\alpha_m=g_m^2/(4\pi)$ characterizes the strength of 
the gauge interactions responsible for the two-loop communication.
The importance of $F/F_S$ derives from the following
inequality \cite{itp96} (see also~\cite{cosmur}):
$(\sqrt F/2000\tev)\geq \sqrt{F/F_S}
(\mslepr/100\gev)/(27.5\sqrt{\nmess})$.\footnote{This inequality is 
%based, in part, on Eqs.~(\ref{inomasses}) and (\ref{scalarmasses}) and is 
valid to within a model-dependent factor of a few.}
For $\mslepr\geq 45\gev$ (the rough LEP limit)
and the two-loop communication value of
$F/F_S=2.5\times10^4$, $\sqrt F\geq 5200\tev$ ($\sqrt F\geq 2600\tev$)
for $\nmess=1$ ($\nmess=4$);
\ie\ even allowing for a factor
of 2 or 3 uncertainty in the approximate lower bound, $\sqrt F$ must lie
near its upper limit from cosmology and the NLSP decay length will be long. 
Previous phenomenological analyses~\cite{ddrt,akkmm,dtw,bbct,bmpz}
have assumed that a model with $F/F_S\sim 1$ is possible,
\footnote{To date, only one explicit model \cite{luty} purports 
to allow $F/F_S\sim 1$.}
in which case the rough bound becomes 
$\sqrt F\geq (73\tev/\sqrt{\nmess})(\mslepr/100\gev)$; 
\ie\ $\sqrt F$ could then be small,
although nothing would prohibit $\sqrt F$
values much larger than this lower limit.
Thus, at the very least, 
it is highly relevant to assess how our ability to discover
supersymmetry changes as a function of $\sqrt F$.

Motivated by the $ee\gam\gam$ event at the Tevatron~\cite{park},
in this Letter we will focus on the ($\nmess=1$) scenario in which
the $\cnone$ (primarily $\wtil B$) is the NLSP.
Phenomenology for the case in which $\sqrt F$ is small 
and $\cnone\to\gamma+G$ decay is prompt has been 
studied~\cite{ddrt,akkmm,dtw,bbct,bmpz}. The case where $\sqrt F$ is very large
and most $\cnone$ decays occur outside the detector has not
been examined; it is equivalent
to conventional supersymmetry phenomenology with the constraints
implied by Eqs.~(\ref{inomasses}) and (\ref{scalarmasses}) among
the sparticle masses. Phenomenology for intermediate $\sqrt F$ values
has also not been considered. Given the uncertainty in $\sqrt F$, 
it is appropriate to consider searching simultaneously 
for a set of signals that are sensitive to different $\sqrt F$ values.

We focus on the D0 detector. 
Events were generated simultaneously
for all \susy\ production mechanisms by modifying 
ISASUGRA/ISAJET~\cite{isajet} to incorporate the GMSB boundary conditions,
and then forcing delayed $\cnone$ decays according to the predicted $c\tau$.
We used the toy calorimeter simulation package ISAPLT. We simulated
calorimetry covering $|\eta|\leq 4$ with a cell size given by
$\Delta R\equiv \Delta\eta\times\Delta\phi=0.1\times 0.0875$ and took the
hadronic (electromagnetic) calorimeter resolution to be
$0.7/\sqrt E$ ($0.15/\sqrt E$). The D0 electromagnetic calorimeter
was simplified to a thin cylinder with radius $r=1$m
and length $-2\leq z\leq +2$m. Also important to our analysis will
be the central outer hadronic D0 calorimeter (OHC), which we approximated
as occupying a hollow solid cylinder defined by $-2\leq z\leq +2$m and 
radial region $2\leq r\leq 2.5{\rm m}$. It is important
that the OHC is segmented into units of size $\Delta R\sim 0.1$ 
and that there are also several layers of inner hadronic calorimeter.

The signals are defined in terms of jets, isolated prompt photons and
isolated photons emerging from $\cnone$ decays occurring in the OHC
(isolated OHC photons). 
A jet is defined by requiring
$|\eta_{\rm jet}|<3.5$ and  $E_T^{\rm jet}>25\gev$ (for  
$\Delta R_{\rm coal.}=0.5$). A photon is a prompt photon candidate
if it emerges immediately or from a $\cnone$ decay that
occurs before the $\cnone$ has reached the electromagnetic calorimeter.
The $(\eta_\gamma,\phi_\gamma)$ of such a photon
is defined by the direction
of the vector pointing from the interaction point to the point at which it
hits the electromagnetic calorimeter.
(This is generally not the same as the direction of the photon's momentum.)
An isolated prompt photon is defined by requiring
$|\eta_\gamma|<1$ and $E_T^\gamma>12\gev$,
with isolation specified by $E_T(\Delta R\leq 0.3)<4\gev$ (summing over 
all other particles in the cone surrounding the photon).
Photons that emerge within the hollow cylinder defined by 
the electromagnetic calorimeter,
but that are not isolated, are merged with hadronic jets as appropriate.
An isolated OHC photon is defined as one that emerges from a
$\cnone\to \gamma G$ decay that occurs within the body of the central OHC
and has $E_T^\gamma>15\gev$, with isolation specified
by $E_T(\Delta R\leq 0.5)<5\gev$ (summing over all other particles
in the cone surrounding the location of the $\cnone$ decay).
For all signals, events are retained only if at least one of several sets 
of reasonable trigger requirements (too
numerous to list here) are satisfied.

We consider three signals:
\begin{trivlist}
\item[ I:] The standard \susy\ signal of jets plus missing energy,
dominant when the $c\tau$ for the $\cnone$ is large. We employ D0 cuts
\cite{dzero}: 
(a) $n({\rm jets})\geq 3$ --- labelled $k=1,2,3$ according to decreasing $E_T$;
(b) no isolated ($E_T^{\rm had.}(\Delta R\leq 0.3)<5\gev$)
leptons with $E_T>15\gev$;
(c) $\etmiss>75\gev$; 
(d) $0.1<\Delta\phi(\etmiss,j_k)<\pi-0.1$ and
$\sqrt{(\Delta\phi(\etmiss,j_1)-\pi)^2+(\Delta\phi(\etmiss,j_2))^2}>0.5$.
The background cross section level has been estimated by D0 to be
$\sigma_B=2.35\pb$. The signal will be deemed observable if: (i)
there are at least 5 signal events; (ii) $\sigma_S/\sigma_B>0.2$; and (iii)
$N_S/\sqrt{N_B}>5$, where $N_S$ and $N_B$ are the numbers of signal
and background events.
\item[ II:] The prompt $2\gamma$ signal, dominant when $c\tau$ for the $\cnone$
is small. Following Ref.~\cite{bbct}, 
we require: (a) at least two isolated prompt photons (as defined above); 
and (b) $\etmiss>\etmin$.
Detection efficiency of 80\% (100\%) is assumed if 
$E_T^\gamma<25\gev$ ($>25\gev$). This signal should be completely
free of background provided $\etmin$ is adjusted appropriately
as a function of luminosity.
Based on the background results of \cite{d0gamgam} (see also
\cite{cdfgamgam}) and eye-ball extrapolations thereof, we estimate
$\etmin=30,50,70\gev$ is sufficient for $L=100\pbi,2\fbi,30\fbi$,
respectively. The signal 
will be deemed observable if there are 5 or more events.
\item[ III:]  A delayed $\gamma$ appearance signal, useful for moderate to large 
$c\tau$ values. The best such signal is highly detector dependent.
As already stated, we shall focus on signals associated with the outer
hadronic calorimeter of the D0 detector. Consider an event in which
a $\cnone\to \gamma G$ decay occurs inside one
of the outer hadronic calorimeter (OHC) cells~\cite{mani}. The $\gamma$ will
deposit all its energy in the cell. By demanding substantial $\gamma$ energy 
and isolation (precise requirements were given earlier)
for this deposit, along with other criteria, backgrounds
can be made small.
\footnote{Our additional criteria will be chosen so that the cross section for
producing an isolated energetic long-lived kaon in association
therewith is small. Further, any such kaon will interact
strongly and is almost certain to be absorbed before reaching the OHC.}
More specifically, we require that the event fall into
one of three classes defined by the following sets of requirements:
%(1) that there is at least one such isolated $\cnone$ decay 
%in an OHC cell in association with three energetic jets and missing energy; 
%(2) that there are two isolated $\cnone$ decays in different OHC
%cells in association with missing energy; or 
%(3) that there is one isolated $\cnone$ decay in an OHC cell in association 
%with one or more isolated prompt photons and missing energy.
%The specific cuts for these three classes of events are:  
\begin{enumerate} 
\item
(a) $n({\rm jets})\geq 3$; 
(b) at least one isolated OHC photon;
(c) $\etmiss>\etmin$. 
\item 
(a) any number of jets;
(b) two or more isolated OHC photons;
(c) $\etmiss>\etmin$. 
\item 
(a) any number of jets;
(b) at least one isolated prompt photon;
(c) at least one isolated OHC photon;
(d) $\etmiss>\etmin$. 
\end{enumerate}
In the absence of the needed
detector-specific study, we have assumed zero background to 1.+2.+3.
for the same $\etmin=30,50,70\gev$
values at $L=100\pbi,2\fbi,30\fbi$ as employed
for the prompt $2\gamma$ signal. 
Observability of the delayed-decay signals is assumed if the number of
events for 1.+2.+3. is 5 or more.

\end{trivlist}

\begin{figure}[ht]
\leavevmode
\epsfxsize=5.5in
\centerline{\epsffile{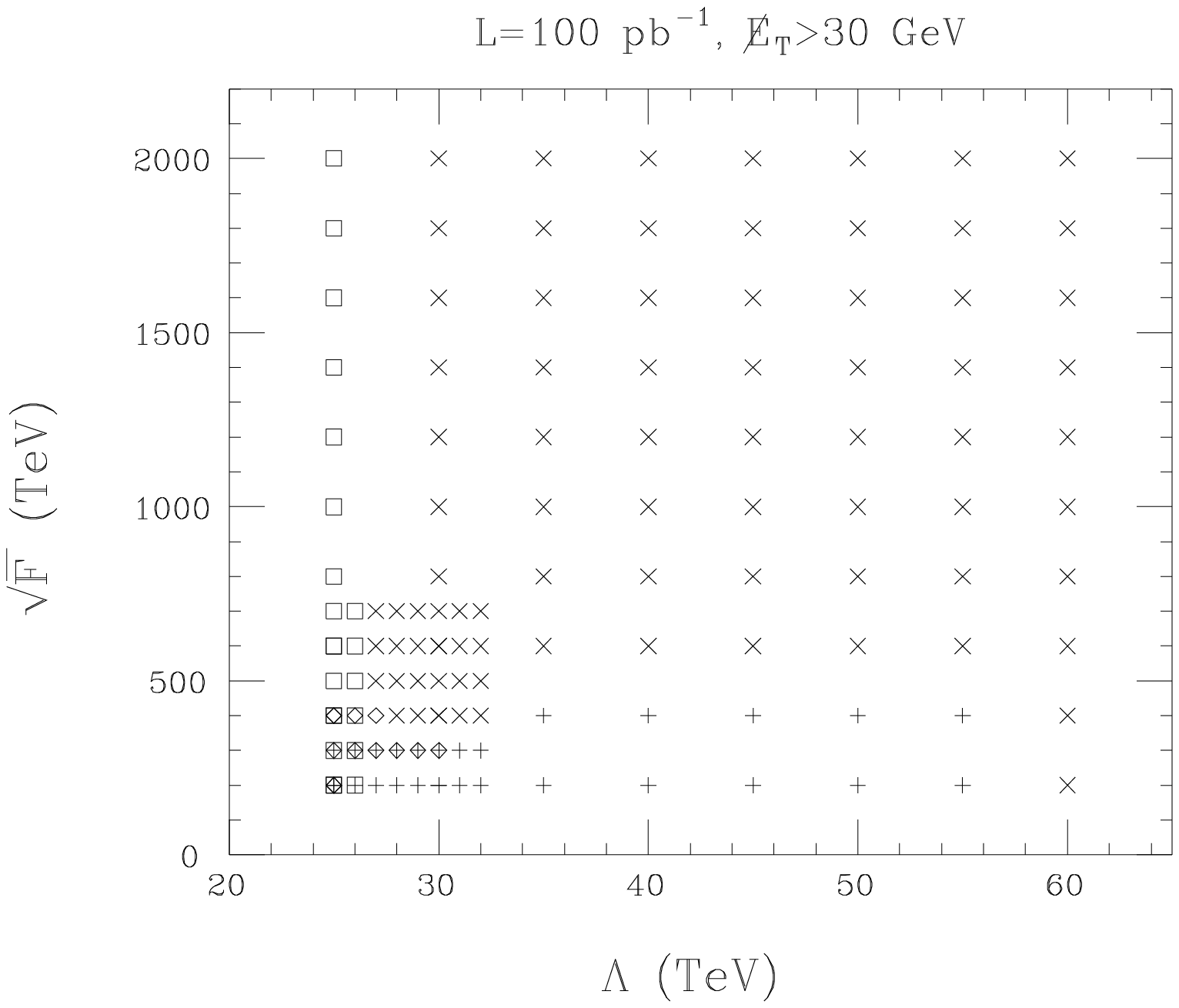}}
\bigskip
\caption{
For $L=100\pbi$ and $\etmin=30\gev$,
we display those points in $(\Lambda,\protect\sqrt F)$
parameter space for which we can observe the following GMSB \susy\ signals: 
$\Box$, jets plus $\etmiss$;
$+$, prompt $2\gamma$;
$\Diamond$, OHC delayed-decay.
At points labelled by $\times$, no signal is observable.}
\label{lumi}
\end{figure}

\begin{figure}[ht]
\leavevmode
\epsfxsize=5.5in
\centerline{\epsffile{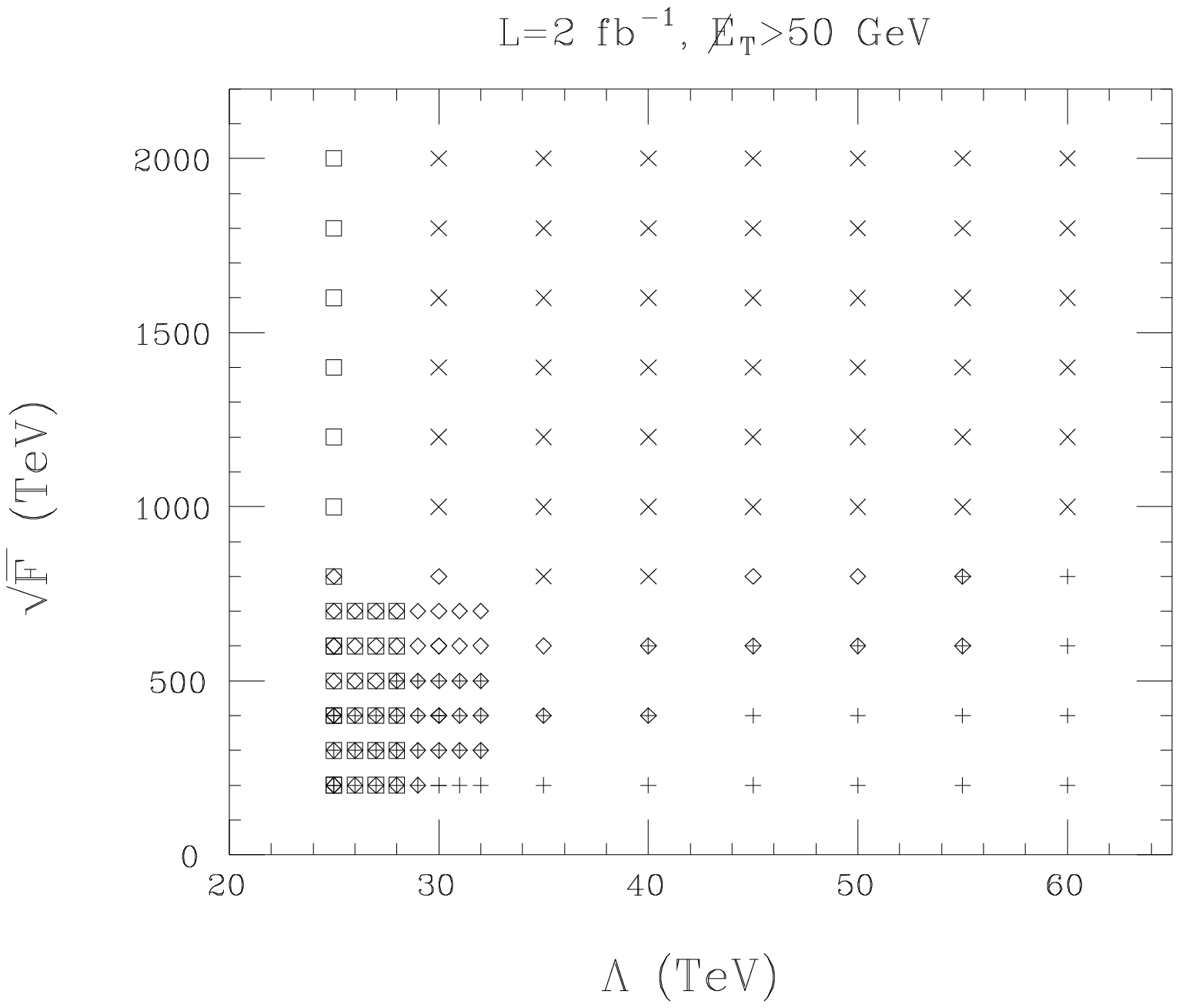}}
\bigskip
\caption{As in Fig.~\protect\ref{lumi}, but for $L=2\fbi$ and $\etmin=50\gev$.}
\label{lumii}
\end{figure}

\begin{figure}[ht]
\leavevmode
\epsfxsize=5.5in
\centerline{\epsffile{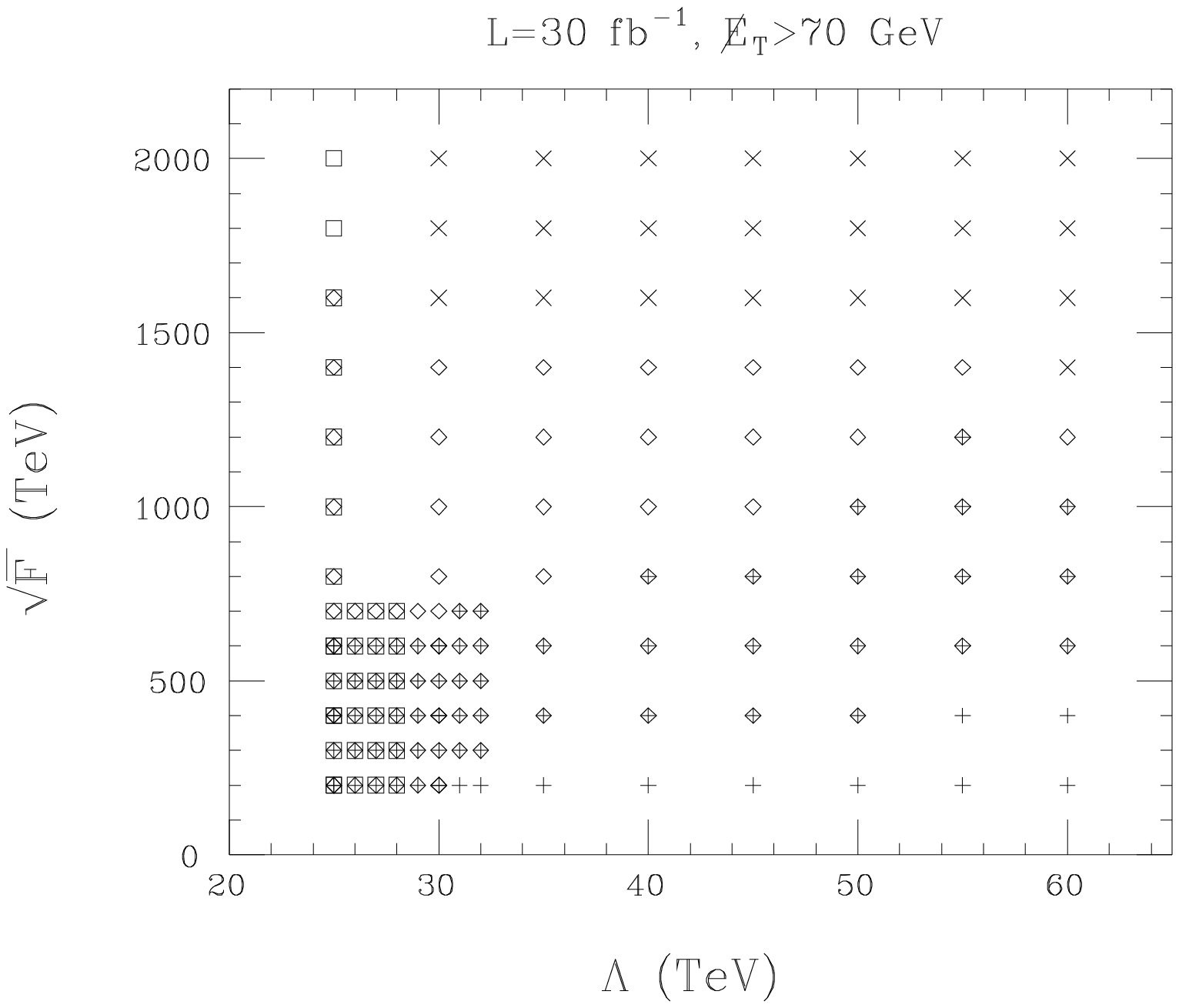}}
\bigskip
\caption{As in Fig.~\protect\ref{lumi}, but for $L=30\fbi$ and $\etmin=70\gev$.}
\label{lumiii}
\end{figure}

The results of our analysis are displayed in Figs.~\ref{lumi},
\ref{lumii} and \ref{lumiii}, for integrated luminosities
$L=100\pbi$, $2\fbi$ and $30\fbi$, respectively.
For $L=100\pbi$, we observe that the prompt $2\gamma$ signal ($+$ points)
allows \susy\ detection out to $\Lambda\sim 55\tev$ (equivalent
to $\mbino\sim 70\gev$) so long
as $\sqrt F\lsim 400\tev$ (\ie\ such that most $\cnone$ decays are prompt).
The standard jets plus $\etmiss$ signature ($\Box$ points) 
is viable only for
small $\Lambda\lsim 26\tev$. The delayed $\cnone$ decay signal(s)
($\Diamond$ points) expand the discovery region 
by only one scan point: $(\sqrt F,\Lambda)=
(400\tev,27\tev)$. But, they are viable at
some points where the previous two signals are observable and would then
give an indication of the decay length of the $\cnone$.

If D0 fails to observe the jets plus $\etmiss$ signal in
their current $L=100\pbi$ data then all the $\Box$ points
in Fig.~\ref{lumi} will be excluded. Further,
recent analyses by D0~\cite{d0gamgam} and CDF~\cite{cdfgamgam} 
place a 95\% confidence level of less than 1 event on the prompt
$2\gamma$ signal for essentially the same cuts and luminosity
that we have considered here, which eliminates all the $+$ points
on this same plot. Thus, only if the stringent cuts we have
employed for the delayed-decay signals could be weakened without
encountering backgrounds would present data allow elimination
of a significant additional portion of parameter space.

For $L=2\fbi$ (the nominal Run II integrated luminosity per detector),
the parameter regions for which the prompt $2\gamma$ and delayed-decay
signals are observable both expand dramatically.
At moderate $\Lambda\lsim 50\tev$,
the delayed-decay signals probe out to larger $\sqrt F$
than does the prompt $2\gamma$ signal, whereas for $\Lambda\gsim 30\tev$
only the prompt $2\gamma$ signal is viable for the smallest
$\sqrt F$ values. However,
there are many points where both types of signal would be detectable.
The region of viability for the
standard jets + $\etmiss$ signal expands only slightly to 
$\Lambda\lsim 28\tev$.

For $L=30\fbi$ (the nominal TeV33 integrated luminosity per detector),
the delayed-decay signals cover more of parameter space than either
of the other two signals, with sensitivity out to $\sqrt F=1400\tev$ 
for all $\Lambda$ values explored. The region of overlap
with the prompt $2\gamma$ signal also expands significantly.
The jets + $\etmiss$ signature is still confined to relatively
small $\Lambda\lsim 28\tev$ values (for any $\sqrt F$).

Of the three delayed-decay signals, 
the first ($\geq 3$ jets + $\geq 1$ OHC $\gamma$)
is observable (\ie\ yields at least five events) for {\it all} 
$(\Lambda,\sqrt F)$ points where the delayed-decay signals are shown
to be viable (the $\Diamond$ points in the figures).  
The second ($\geq 2$ OHC $\gamma$'s) yields $\geq 5$ events 
for very few points.
The third ($\geq 1$ prompt $\gamma$ + $\geq 1$ OHC $\gamma$)
yields $\geq 5$ events for the lower (roughly, half) $\sqrt F$ portion of the
$\Diamond$ regions of the figures.
If simulations eventually
show that the first signal is background-free for smaller $\etmin$ than
employed here (as we are hopeful will be
the case), the portion of parameter space
for which it is viable expands; \eg\ at $L=30\fbi$ and $\etmin=30\gev$
sensitivity would extend to $\sqrt F=1800\tev$ for most of the $\Lambda$
values shown.

We now briefly outline other possible delayed-decay signatures.
\begin{itemize}
\item
Timing --- Since the $\cnone$ moves with $v/c$ substantially
less than 1, a late energy deposit would be a signal. In Run I,
only CDF had any timing information for hadronic calorimeters, but
some timing information will be available at both CDF and D0 
for Run II~\cite{features}.
\item
Impact parameter --- If the impact parameter of a photon appearing
in the electromagnetic calorimeter could be shown to be non-zero,
this would also constitute a signal for a delayed decay.  The extent to which
the resolution for such an impact parameter measurement by the
CDF and D0 detectors is adequate to establish a signal
is under study. The LHC detectors have sufficiently better resolution
that the non-zero impact parameter signal should be quite viable there.
\item
Timing and directional information outside the muon chambers ---
For large $\sqrt F$, many $\cnone$ decays will occur outside
the muon chambers.  A multi-layer tracking array supplement to the
cosmic ray veto would allow detection of such decays and provide
a dramatic signal. Such an array would be useful for any
delayed $\cnone$ decay scenario.
\end{itemize}
Sensitivity to GMSB models with substantial $\cnone$ decay length will
be maximized by implementing all possible signals simultaneously.
One can also envision
construction of a special purpose detector for delayed decays~\cite{itp96}.
One such proposal for the LHC has appeared~\cite{orito}.

If delayed decay signatures are seen, the next goal will be to determine
the values of $\sqrt F$ and $\Lambda$. The least model-dependent 
way of obtaining $\sqrt F$ is to determine $c\tau$
and $\mcnone$ [see Eq.~(\ref{ctauform})]; $\mcnone$ then
implies via Eq.~(\ref{inomasses}) a value for $\Lambda$.
To determine $c\tau$ requires knowledge of the distribution of $\cnone$
decays.  If both the prompt $2\gamma$ and OHC signals are seen,
their relative rates will be a measure of $c\tau$.
One way to measure $\mcnone$ would
be to determine the time of flight to the OHC cell (requiring a 
measurement of the time of the OHC energy deposit) and correlate
this with the energy of the deposit (\ie\ the energy of the photon)
which would reflect the energy of the decaying $\cnone$.
Substantial statistics will be required for both of these analyses.

In conclusion, we have argued that it is very natural (indeed,
preferred in most existing models)
for the general $(\Lambda,\sqrt F)$ parameters of a gauge-mediated
supersymmetry breaking model to be such that there is a large probability for
the lightest neutralino to decay (to a photon plus goldstino)
a substantial distance from the interaction point.
We have shown that the portion of the parameter space for which
a signal for supersymmetry can be seen at the Tevatron
is greatly expanded by employing signals sensitive to such delayed decays.
The CDF and D0 detector groups should give
detailed consideration to clarifying and refining such signatures.

\vspace{2cm}
\centerline{\bf Acknowledgments}
\vspace{.5cm}
This work was supported in part by the U.S. Department of Energy under
grant No. DE-FG03-91ER40674, and
by the Davis Institute for High Energy Physics. We are grateful 
to B. Dobrescu, S. Kuhlmann, S. Mani, H. Murayama, L. Nodulman, 
S. Thomas, A. White and A. Wicklund for helpful conversations.

\clearpage

\phantom{skip}
\vfill

\vfill

\begin{thebibliography}{99}



\bibitem{ddrt} 
S. Dimopoulos, M. Dine, S. Raby
and S. Thomas, \prlj{76}{3494}{1996}.

\bibitem{akkmm}
S. Ambrosanio, G. Kane, G. Kribs, S. Martin and S. Mrenna, 
\prlj{76}{3498}{1996}.

\bibitem{dtw}
S. Dimopoulos, S. Thomas and J.D. Wells, \prdj{54}{3283}{1996}.

\bibitem{bbct}
H. Baer, M. Brhlik, C.-H. Chen and X. Tata, \prdj{55}{4463}{1997}.

\bibitem{bmpz}
J. Bagger, K. Matchev, D. Pierce and R.-J. Zhang,
\prlj{78}{1002}{1997}.

\bibitem{basicgmsb} See, for example, 
M. Dine, A. Nelson and Y. Shirman, 
\prdj{51}{1362}{1995}; M. Dine, A. Nelson, Y. Nir and Y. 
Shirman, \prdj{53}{2658}{1996}; 
and references to earlier work therein.

\bibitem{fayet} P. Fayet, \plbj{70}{461}{1977}; \plbj{86}{272}{1979};
\plbj{175}{471}{1986}.

\bibitem{others} N. Cabibbo, G.R. Farrar and L. Maiani, \plbj{105}{155}{1981};
M.K. Gaillard, L. Hall and I. Hinchliffe, \plbj{116}{279}{1982};
J. Ellis and J.S. Hagelin, \plbj{122}{303}{1983}; D.A. Dicus, S. Nandi
and J. Woodside, \plbj{258}{231}{1991}.

\bibitem{martin} S. Martin, hep-ph/9608224.

\bibitem{coslimit} H. Pagels and J.R. Primack, \prlj{48}{223}{1982}; 
T. Moroi, H. Murayama and M. Yamaguchi, \plbj{303}{289}{1993};
S. Borgani, A. Masiero and M. Yamaguchi, \plbj{386}{189}{1996}.


\bibitem{cosmur} A. de Gouvea, T. Moroi and H. Murayama, hep-ph/9701244.

%\bibitem{murayama} K. Intriligator and S. Thomas, \npbj{473}{121}{1996};
%E. Poppitz and S. Trivedi, \prdj{55}{5508}{1997}, hep-ph/9701286;
%N. Arkani-Hamed, J. March-Russell and H. Murayama,
%hep-ph/9701286; H. Murayama, hep-ph/9705271.
%See also, G. Dvali and M. Shifman, hep-ph/9612490.

\bibitem{itp96} J.F. Gunion, Proc. ITP Conference on Future High-Energy
Colliders, Santa Barbara, CA, 21-25 Oct. 1996, hep-ph/9704349.

\bibitem{luty} M. Luty, hep-ph/9706554.

\bibitem{park}
S. Park (CDF Collaboration), ``Search for New Phenomena in CDF'', 10th
Topical Workshop on Proton--Antiproton Collider Physics, eds. R. Raja
and J. Yoh, AIP Press, 1995.


\bibitem{dzero}  S. Abachi \etal\ (D0 Collaboration), \prlj{75}{618}{1995}.

\bibitem{d0gamgam} S. Abachi \etal\ (D0 Collaboration), \prlj{78}{2070}{1997}.

\bibitem{cdfgamgam} D. Toback (CDF Collaboration), FERMILAB-Conf-96/240-E.

\bibitem{mani} The general idea of the OHC
signal was developed in discussions with S. Mani.

\bibitem{isajet} F. Paige and S. Protopopescu, in {\it Supercollider Physics},
p. 41, ed. D. Soper (World Scientific, 1986); H. Baer, F. Paige, S.
Protopopescu and X. Tata, hep-ph/9305342.

\bibitem{features} The rough detector features noted were gleaned
from discussions with S. Kuhlmann (CDF), L. Nodulman (CDF), 
A.B. Wicklund (CDF), and S. Mani (D0).

\bibitem{orito} K. Maki and S. Orito, hep-ph/9706382.

\end{thebibliography}
\end{document}